\documentclass[psfig,epsf]{article}
\usepackage{epsfig} \usepackage{amssymb} \usepackage{amsfonts}

\def\beq{\begin{equation}}
\def\eeq{\end{equation}}
\def\bea{\begin{eqnarray}}
\def\eea{\end{eqnarray}}
\def\bem{\begin{math}}
\def\eem{\end{math}}
\def\bit{\begin{itemize}}
\def\eit{\end{itemize}}
\def\bla{\begin{flushright}}
\def\ela{\end{flushright}}
\def\qq2{$Q^2$}               % Q2
\def\aa1{$A_1(x,Q^2)$}        % A1
\def\ff1{$F_1(x,Q^2)$}        % F1
\def\gg1{$g_1(x,Q^2)$}        % g1
\newpage
\newcommand{\df}[2]{\mbox{$\frac{#1}{#2}$}}

\newcommand{\la}{\lambda}
\newcommand{\De}{\Delta}
\newcommand{\rd}{{\rm d}}

\begin{document}
%---------------------------------------------------------------------

%                          Title
\begin{center}
{\Large \bf The Gegenbauer Polynomial Technique:\\ 
the evaluation of 
%a class of  Feynman
complicated
%diagrams
Feynman integrals
} \\

\vspace{4mm}

%                      author/address
A.V.Kotikov\\
Particle Physics Laboratory,
%\\ 
Joint Institute for Nuclear Research \\
141980 Dubna, Russia.\\
\end{center}

%                        Abstract
\begin{abstract}

We discuss a progress in calculation of Feynman integrals which has been
done with help of the Gegenbauer Polynomial Technique and 
demonstrate the results for 
most complicated parts of ${\rm O}(1/N^3)$ contributions to 
%the large-$N$ 
critical exponents 
%$\eta$ 
of
%the non-linear $\sigma$-model, and hence 
$\phi^4$-theory,
for any spacetime dimensionality $D$.\\
%a class of two-point two-loop diagrams.\\

\end{abstract}

\section{Basic Formulae}

Fifteen years ago
the method based on the expansion of propagators in Gegenbauer series
(see \cite{1.1}) has been introduced in \cite{2,2.1}.
 One has shown  \cite{2,ChSm} that by this
method the analytical evaluation of counterterms in the minimal
subtraction scheme at the 4-loop level in any model and for any
composite operator was indeed possible. The Gegenbauer Polynomial (GP)
 technique has been applied
successfully for propagator-type Feynman diagrams (FD) in many calculations 
(see \cite{2,2.1}).
In the present paper we consider
{\it a 
development of the GP technique} (obtained in \cite{4}) and the application
of the results in \cite{BrKo}.

Throughout the paper we use the following notation. The use of dimensional
regularization is assumed. All the calculations are performed in the
%coordinate 
space
 of dimension $D=4-2\varepsilon$.
Note that contrary to \cite{2} we
analyze FD directly in  momentum $x$-space which allows
us to avoid the appearance of Bessel functions.
%It is possible 
Because we consider here only
%in the case of 
propagator-type massless FD, we know
%because 
their
dependence on a single external momentum 
%(in $p$-space) or on a single external coordinate is power-like and known 
beforehand. The point of
interest is the coefficient function $C_f$, which depends on
$D=4-2\varepsilon$ and is a Laurent series in $\varepsilon $. \\

{\bf 1.}
First of all,
%Here
we present useful formulae to use of Gegenbauer polynomials.
% and traceless products.
Following \cite{2,2.1}, $D$-space integration can
be represented in the form
$$
d^Dx~=~ \frac{1}{2} S_{D-1}(x^2)^{\lambda} dx^2 d\hat{x}~~~~
(\lambda = D/2-1),
$$
where $
%\bf{x}^2=\vec{x}^2,
{}~\hat{x}=\vec{x}/\sqrt{
%\vec{x}
x^2}$, 
%$\lambda = D/2-1=1-\varepsilon$
and
$S_{D-1}=2 {\pi}^{\lambda+1}/\Gamma(\lambda+1)$ is the surface of the
unit hypersphere in $R_D$. The Gegenbauer polynomials
$C_n^{\delta}(t)$ are defined as \cite{1.1,2.1}
\begin{eqnarray}
 (1-2rt + r^2)^{-\delta} ~=~ \sum^{\infty}_{n=0}~C_n^{\delta}(t)r^n
{}~~~~(r \leq 1),~~~~ C_{n}^{\delta}(1) ~=~ \frac{ \Gamma(n+2\delta)}
{n!\Gamma(2\delta)},
  \label{A1}
\end{eqnarray}
whence the expansion for the propagator is:
\begin{eqnarray}
\frac{1}{(x_1-x_2)^{2\delta}} ~=~
\sum^{\infty}_{n=0} ~C_n^{\delta}(\hat{x_1}\hat{x_2})~
\Biggl[
\frac{{(x^2_1)}^{n/2}}{{(x^2_2)}^{n/2 + \delta}} \Theta(x_2^2 - x_1^2)
{}~+~
\Bigl(x_1^2 \longleftrightarrow x_2^2  \Bigr)  \Biggr],
  \label{A2}
\end{eqnarray}
where
\[ \Theta(y)~=~ \left\{
\begin{array}{rl}
1, & \mbox{ if }y \geq 0 \\
0, & \mbox{ if }y  <   0
\end{array} \right.  \]

Orthogonality of the Gegenbauer polynomials $C_n^{\lambda}(x)$
%having the index $\lambda$
is expressed by the equation (see \cite{2})
\begin{eqnarray}
\int ~C_n^{\lambda}(\hat{x_1}\hat{x_2})~
 \, C_m^{\lambda}(\hat{x_2}\hat{x_3}) \, d\hat{x_2}
~=~\frac{\lambda}{n+\lambda}\, \delta^m_n
 ~C_n^{\lambda}(\hat{x_1}\hat{x_3}),~
  \label{A3}
\end{eqnarray}
where $\delta^m_n$ is the Kronecker symbol.

%To our study
The following formulae are useful (see \cite{2,3}):
\begin{eqnarray}
%\z
C_n^{\delta}(x)&=&
\sum_{p \geq 0} \frac{(2x)^{n-2p}(-1)^p \Gamma(n-p+\delta)}
{(n-2p)!p!\Gamma(\delta)}
%\label{A4}   \\\z
{}~~~
{}~ \mbox{ and }~
\nonumber   \\ 
%\z
\frac{(2x)^{n}}{n!} &=& \sum_{p \geq 0}
C_{n-2p}^{\delta}(x) \frac{(n-2p+\delta)\Gamma(\delta)}
{p!\Gamma(n-p+\delta+1)}  \label{A4}
%\\ \z   C_{n}^{\delta}(1) ~=~ \frac{ \Gamma(n+2\delta)}
%{n!\Gamma(2\delta)}  \label{A6}
\end{eqnarray}

Substituting the latter equation from (\ref{A4}) for $\delta = \lambda$ to
the first one, we have the following equation after the separate
analyses at odd and even $n$:
\begin{eqnarray}
C_n^{\delta}(x)~=~\sum_{k=0}^{[n/2]}
C_{n-2p}^{\lambda}(x) \frac{(n-2k+\lambda) \Gamma(\lambda)}
{k!\Gamma(\delta)}  \frac{ \Gamma(n+\delta-k)\Gamma(k+\delta-\lambda)}
{\Gamma(n+\lambda+1-k)\Gamma(\delta-\lambda)}  \label{A7}
\end{eqnarray}

%This equation is an example of the
%%one from
%%basical
%%more used
%equations used in our analysis.
%Note that alternatively to (\ref{A7}) we can use  the expansion of
%$(x_1-x_2)^{-2\delta}$ similar to (\ref{A3}) but directly in a series
%of GP $C_n^{\lambda}(\hat x_1 \cdot \hat x_2)$ (see \cite{2.1}).\\

{\bf 2.}
Following \cite{2,3} we introduce the traceless product (TP)
$x^{\mu_1...\mu_n}$ connected with the usual product
$x^{\mu_1}...x^{\mu_n}$ by the following equations 
%(see \cite{2,3})
\begin{eqnarray}
%\z 
x^{\mu_1...\mu_n}&=&\hat{S}
\sum_{p \geq 0} \frac{n!(-1)^{p} \Gamma(n-p+\lambda)}
{2^{2p} p! (n-2p)!\Gamma(n+\lambda)}~
g^{\mu_1\mu_2}...g^{\mu_{2p-1}\mu_{2p}}~x^{2p}~x^{\mu_{2p+1}}...x^{\mu_n}
%\label{B1}   \\
               \nonumber  \\
& & \nonumber  \\
%\z 
x^{\mu_1}...x^{\mu_n}&=&\hat{S}
\sum_{p \geq 0} \frac{n! \Gamma(n-2p+\lambda+1)}
{(2)^{2p} p! (n-2p)!\Gamma(n-p+\lambda+1)}~
g^{\mu_1\mu_2}...g^{\mu_{2p-1}\mu_{2p}}~x^{2p}~x^{\mu_{2p+1}...\mu_n}
\label{B1}
\end{eqnarray}
%where hereafter $\lambda \equiv D/2-1$.

Comparing Eqs.(\ref{A4})
%, (\ref{A6})
and (\ref{B1}), we obtain the
following relations between TP and GP
\begin{eqnarray}
z^{\mu_1...\mu_n}~x^{\mu_1...\mu_n}~=~ \frac{n! \Gamma(\lambda)}
{2^n \Gamma(n+\lambda)}~C_n^{\lambda}(\hat{x}\hat{z})~{(x^2 z^2)}^{n/2},~
%~~\mbox{ and, hence, }~~
~ x^{\mu_1...\mu_n}~x^{\mu_1...\mu_n}~=~ 
\frac{\Gamma(n+2\lambda) \Gamma(\lambda)}
{2^n \Gamma(2\lambda)\Gamma(n+\lambda)}~x^{2n}
  \label{B2}
\end{eqnarray}
We give also the simple but quite useful conditions:
\begin{eqnarray}
z^{\mu_1...\mu_n}~x^{\mu_1...\mu_n}~=~
z^{\mu_1}...\,z^{\mu_n}~x^{\mu_1...\mu_n}~=~
z^{\mu_1...\mu_n}x^{\mu_1}...\,x^{\mu_n},
  \label{B3}
\end{eqnarray}
which
 follow immediately from the TP definition:
$g^{\mu_i\mu_j}~x^{\mu_1...\mu_i...\mu_j...\mu_n}~=~ 0$.

The use of the TP $x^{\mu_1...\mu_n}$ makes it possible to ignore
terms of the type $g^{\mu_i\mu_j}$ that arise upon integration: they can
be readily recovered from the general structure of the TP. Therefore,
in the process of integration it is only necessary to follow the
coefficient of the leading term $x^{\mu_1}...x^{\mu_n}$.
The rules to integrate FD containing TP can be found, for example in
\cite{3,3.1,3.2}. For a loop we have 
(hereafter $Dx \equiv (d^Dx)/(2\pi)^D$)
\footnote{The Eq.(\ref{4}) has been used in \cite{3,3.1,3.2,KazKo} for 
calculations of the moments of structure functions of deep inelastic 
scattering.}:
\begin{eqnarray}
%\z
%~~
\int Dx \frac{ x^{\mu_1...\mu_n}}{x^{2\alpha}(x-y)^{2\beta}}
{}~=~
\frac{1}{(4\pi)^{D/2}}~
\frac{y^{\mu_1...\mu_n}}{y^{2(\alpha + \beta - \lambda -1)}}
{}~ A^{n,0}(\alpha, \beta),
  \label{4}
\end{eqnarray}\\
where
$$A^{n,m}(\alpha,\beta)~=~
\frac{a_n(\alpha)a_m(\beta)}{a_{n+m}(\alpha+\beta-\lambda-1)}
~~\mbox{ and }~~
a_n(\alpha)~=~\frac{\Gamma (D/2-\alpha +n )}{\Gamma(\alpha)}
$$

Note that in our analysis it is necessary to consider more
complicate cases of integration, when the integrand  contains
$\Theta$ functions. Indeed, using the Eqs.(\ref{A2}) and (\ref{B2}),
we can represent
the propagator $(x_1-x_2)^{-2\lambda}$ into the following form
\footnote{In the case of the propagator $(x_1-x_2)^{-2\delta}$ with
$\delta \neq \lambda$ we should use also Eq.(\ref{A7}).}:
\begin{eqnarray}
\frac{1}{(x_1-x_2)^{2\lambda}} ~=~
\sum^{\infty}_{n=0} ~\frac{2^n \Gamma(n+\lambda)}{n!\Gamma(\lambda)}
{}~x_1^{\mu_1...\mu_n}~x_2^{\mu_1...\mu_n}~
\Biggl[
\frac{1}{x_2^{2(\lambda +n)}} \Theta(x_2^2 - x_1^2)
{}~+~
\Bigl(x_1^2 \longleftrightarrow x_2^2  \Bigr)  \Biggr]
  \label{B6}
\end{eqnarray}
%This equation was already applied in Section 3.

Using the GP properties from  previous subsection
and the connection (\ref{B2})
between GP and TP,
we obtain the  rules for calculating
FD with the $\Theta$-terms and TP.\\

{\bf 3.} {\it The rules}
have the following form:
\begin{eqnarray}
%\z
%~~~~~\equiv ~
\int Dx \frac{x^{\mu_1...\mu_n}}{x^{2\alpha}(x-y)^{2\beta}}
~\Theta(x^2-y^2)
%\nonumber \\ \z
&=&
\frac{1}{(4\pi)^{D/2}}~
\frac{y^{\mu_1...\mu_n}}{y^{2(\alpha + \beta - \lambda -1)}}
~ \sum_{m=0}^{\infty}
\frac{B(m,n|\beta,\lambda)}{m+\alpha + \beta - \lambda -1}~ \nonumber \\
%\z 
&\stackrel{(\beta =
  \lambda)}{=}&
 \frac{1}{(4\pi)^{D/2}}~\frac{y^{\mu_1...\mu_n}}{y^{2(\alpha  -1)}}
{}~\frac{1}{\Gamma(\lambda)}
\frac{1}{(\alpha -1)(n+\lambda)}  \label{2} \\
%\end{eqnarray}
& & \nonumber  \\
%\begin{eqnarray}
%\z
%~~~\equiv ~
\int Dx \frac{ x^{\mu_1...\mu_n}}{x^{2\alpha}(x-y)^{2\beta}}
~\Theta(y^2-x^2)
% \nonumber \\ \z
 &=&
\frac{1}{(4\pi)^{D/2}}~
\frac{y^{\mu_1...\mu_n}}{y^{2(\alpha + \beta - \lambda -1)}}
{}~ \sum_{m=0}^{\infty}
\frac{B(m,n|\beta,\lambda)}{m+n- \alpha + \lambda +1}~  \nonumber \\
%\z 
&\stackrel{(\beta =
  \lambda)}{=}&
 \frac{1}{(4\pi)^{D/2}}~\frac{y^{\mu_1...\mu_n}}{y^{2(\alpha -1)}}
{}~\frac{1}{\Gamma(\lambda)}
\frac{1}{(n+\lambda +1- \alpha )(n+\lambda)}  \label{3}
\end{eqnarray}\\
where
$$B(m,n|\beta,\lambda)~=~\frac{\Gamma(m+\beta +n)}
{m! \Gamma(m+n+1+\lambda) \Gamma(\beta)}~
\frac{\Gamma(m+\beta - \lambda)}{\Gamma(\beta - \lambda)}~~
%\mbox{ and }~~ \lambda=D/2-1
$$

The sum of above diagrams does not contain $\Theta$-terms
and should reproduce Eq.(\ref{4}).
To compare the r.h.s. of Eqs.(\ref{2},\ref{3}) and the r.h.s. of
Eq.(\ref{4}) we use the
transformation of $_3F_2$-hypergeometric function with unit argument
${}_3F_2(a,b,c;e,b+1;1)$
(see \cite{5}):
\begin{eqnarray}
%\z
 \sum_{k=0}^{\infty} \frac{\Gamma(k+a)\Gamma(k+c)}
{k!\Gamma(k+f)} \frac{1}{k+b}
{}&=&\frac{\Gamma (a)\Gamma(1-a)\Gamma(b)\Gamma(c-b)}
{\Gamma(f-b)\Gamma(1+b-a)} \nonumber  \\
%\z
&-&  \frac{\Gamma(1-a)\Gamma(a)}{\Gamma(f-c)\Gamma(1+c-f)}\,
%\cdot~
 \sum_{k=0}^{\infty} \frac{\Gamma(k+c-f+1)\Gamma(k+c)}
{k!\Gamma(k+1+c-a)} \frac{1}{k+c-b}
  \label{8.2}
\end{eqnarray}
This is the case (when $k=m, b=\alpha+\beta-\lambda-1 , c=n+\beta$) to
compare Eq.(\ref{4}) and the sum of Eqs.(\ref{2},\ref{3}).

Analogously to Eqs.(\ref{2}) and (\ref{3}) we have more
complicate cases:
%\footnote{The full set of rules will be presented in  \cite{4}.}:
\begin{eqnarray}
& &\int Dx \frac { x^{\mu_1...\mu_n}}{x^{2\alpha}(x-y)^{2\beta}}
\Theta(x^2-z^2)
      ~=~ \frac{1}{(4\pi)^{D/2}}~
y^{\mu_1...\mu_n}
\Biggr[
\frac{\Theta (y^2 -z^2)}{y^{2(\alpha + \beta - \lambda -1)}}
{}~ A^{n,0}(\alpha,\beta)
 \nonumber \\ 
& & + ~
 \sum_{m=0}^{\infty}
~\frac{B(m,n|\beta,\lambda)}{z^{2(\alpha + \beta - \lambda -1)}}
 ~ \biggl( {\Bigl(\frac{y^2}{z^2}\Bigr)}^{m}
\frac{\Theta (z^2 -y^2)}{m+\alpha + \beta - \lambda -1}~-~
{\Bigl(\frac{z^2}{y^2}\Bigr)}^{m+ \beta +n}
\frac{\Theta (y^2 -z^2)}{m-\alpha +n+1+ \lambda } \biggr)
 \Biggr]
                         ~ \nonumber \\
& & \hspace{2cm}
\stackrel{(\beta =
  \lambda)}{=}
 \frac{1}{(4\pi)^{D/2}}~ \frac{1}{\Gamma(\lambda)}~y^{\mu_1...\mu_n}~
\Biggl[
\frac{1}{y^{2(\alpha -1)}}
\frac{\Theta (y^2-z^2)}{(\alpha -1)(n+\lambda +1 -\alpha)}  \nonumber \\
& &\hspace{2.5cm} 
+~ \frac{1}{z^{2(\alpha -1)}} \frac{1}{n+\lambda }
\biggl(
\frac{\Theta (z^2-y^2)}{\alpha -1}~-~
{\Bigl(\frac{z^2}{y^2}\Bigr)}^{n+ \lambda }
\frac{\Theta (y^2-z^2)}{n+1+ \lambda -\alpha } \biggr)
%]
 \Biggr]  \label{5}\\
& &  \nonumber  \\
& & \int Dx \frac { x^{\mu_1...\mu_n}}{x^{2\alpha}(x-y)^{2\beta}}
\Theta(z^2-y^2)
{}=~ \frac{1}{(4\pi)^{D/2}}~
y^{\mu_1...\mu_n}~
\Biggl[
\frac{\Theta (z^2 -y^2)}{y^{2(\alpha + \beta - \lambda -1)}}
{}~A^{n,0}(\alpha,\beta)
 \nonumber \\ 
& & - ~  \sum_{m=0}^{\infty}
{}~\frac{B(m,n|\beta,\lambda)}{z^{2(\alpha + \beta - \lambda -1)}}
{}~\biggl( {\Bigl(\frac{y^2}{z^2}\Bigr)}^{m}
 \frac{\Theta (z^2 -y^2)}{m+\alpha + \beta - \lambda -1}~-~
{\Bigl(\frac{z^2}{y^2}\Bigr)}^{m+ \beta +n}
\frac{\Theta (y^2 -z^2)}{m-\alpha +n+1+ \lambda } \biggr)
 \Biggr]
{}~\nonumber \\
& & \hspace{2cm} 
\stackrel{(\beta =
  \lambda)}{=}
 \frac{1}{(4\pi)^{D/2}}~ \frac{1}{\Gamma(\lambda)}~y^{\mu_1...\mu_n}~
\Biggl[
\frac{1}{y^{2(\alpha -1)}}
\frac{\Theta (z^2-y^2)}{(\alpha -1)(n+\lambda +1 -\alpha)}  \nonumber \\
& & \hspace{2.5cm}
 -~ \frac{1}{z^{2(\alpha -1)}} \frac{1}{n+\lambda }
\biggl(
\frac{\Theta (z^2-y^2)}{\alpha -1}~-~
{\Bigl(\frac{z^2}{y^2}\Bigr)}^{n+ \lambda }
\frac{\Theta (y^2-z^2)}{n+1+ \lambda -\alpha } \biggr)
 \Biggr]  \label{6}
\end{eqnarray}\\
One can easily see that the sum of the above diagrams lead to
results identical to (\ref{4}).\\

%\section{The evaluation of a class of Feynman diagrams}

\section{Calculation of complicated FD}

% {\bf 2.} 
{\it The
aim of this section} is to demonstrate the result of \cite{4} for
%study  
a class of master
two-loop diagrams containing the vertex with two propagators having index
1 or $\lambda$.

%{\bf 1.}
Consider 
%in the $x$-space 
the following general diagram
$$ \int  \frac {Dx Dy}{y^{2\alpha}(z-y)^{2t}(z-x)^{2\beta}x^{2\gamma}
(x-y)^{2s}}
{}~\equiv ~J(\alpha,t, \beta, \gamma, s)$$
and restrict ourselves to the FD
$A(\alpha, \beta, \gamma ) ~=~ J(\alpha,\lambda, \beta, \gamma,
\lambda)$,
which is the one
of FD of interest for us here. It is easily
shown
(see \cite{VPH,3.2,4})
that ($\sigma = 3+ \lambda -(\alpha + \beta + \gamma )$)
\begin{eqnarray}
C_f[A(\alpha,\beta, \gamma)]~=~C_f[A(\alpha,\sigma, \gamma)]~=~
C_f[J(\gamma,\lambda, \lambda,\sigma, \alpha)]~=~
C_f[J(\sigma,\gamma,\lambda, \lambda,\beta)],   \label{7}
\end{eqnarray}
%{\bf 2.}
 Doing Fourier transformation of both: the diagram $A(\alpha,\beta,\gamma)$
%from eq.(\ref{8})
and its solution
in the form $C_f[A(\alpha, \beta, \gamma)](z^2)^{-\tilde \sigma}$,
where hereafter
$\tilde t = \lambda +1-t,~t= \{\alpha,\beta, \gamma, \sigma , ...\}$, and
considering the new
diagram as one in the momentum
$x$-space 
%(i.e. making the dual transformation)
we obtain the relation
\begin{eqnarray} 
%\z
C_f[A(\alpha,\beta, \gamma)] &=&
%&\stackrel{f}{=}&
\frac{a_0^2(\lambda)a_0(\alpha)a_0(\beta)a_0(\gamma)}{a_0(\delta)}~
C_f[J(\tilde \alpha,1,\tilde \beta,\tilde \gamma ,1)] 
\nonumber\\ 
%\z
%& \stackrel{d}{=}&
&=&
a_0^2(\lambda)a_0(\alpha)a_0(\beta)a_0(\gamma)a_0(\sigma)~
C_f[J(\tilde \beta,1,\tilde \alpha,\tilde \gamma,1)]~  \label{9}
\end{eqnarray}
between the diagram, which contains the vertex
with two propagators
having the index $\lambda$, and the similar diagram containing the vertex
with two 
propagators having the index 1.

%3.
Repeating the manipulations of e would(see \cite{VPH,3.2,4})
%subsection 1 
we can obtain the
following relations:
\begin{eqnarray}
%\z
C_f[J(\tilde \beta,1,\tilde \alpha,\tilde \gamma,1)]=
C_f[J(\tilde \beta,1,\tilde \sigma,\tilde \gamma,1)]
%\nonumber \\ \z
=C_f[J(1,1,\tilde \gamma ,\tilde \sigma,\tilde \alpha )]=
C_f[J(\tilde \sigma,1,1,\tilde \gamma,\tilde \beta)]  \label{10}
\end{eqnarray}
Thus, we have obtained the relations between all diagrams from the class
introduced in the beginning of this section. Hence, it is necessary
to find the solution for one of them. We prefer to analyze the
diagram $A(\alpha,\beta,\gamma)$, that is the content of the next 
subsection.\\

%\section{The calculations}
{\bf 1.}
{\it We calculate the diagram} $A(\alpha, \beta, \gamma)$ by the
following way\footnote{The symbol $\stackrel{(n)}{=}$ marks the fact
  that the equation $(n)$ is used on this step.}:
\begin{eqnarray}
%\z 
& & A(\alpha, \beta, \gamma) \stackrel{(\ref{B6})}{=}
% \mbox{ from Eq.(\ref{B6}) }=~
\sum_{n=0}^{\infty} \frac{2^n \Gamma(n+\lambda)}{n! \Gamma(\lambda)}
\int Dx Dy \frac { z^{\mu_1...\mu_n}}{x^{2\gamma}(z-x)^{2\beta}}
 \frac { y^{\mu_1...\mu_n}}{y^{2\alpha}(x-y)^{2\lambda}}
\Bigl[\frac{\Theta(z^2-y^2)}{z^{2(n+\lambda)}}~+~
\frac{\Theta(y^2-z^2)}{y^{2(n+\lambda)}} \Bigr]~
%\mbox{ from eq.(\ref{5},\ref{6}) }=~
 \nonumber \\
& & 
%\hspace{0.5cm} 
 \stackrel{(\ref{5},\ref{6})}{=}
 \frac{1}{(4\pi)^{D/2}}~ \frac{1}{\Gamma(\lambda)}~\frac{1}{\alpha -1}
\sum_{n=0}^{\infty} \frac{2^n \Gamma(n+\lambda)}{n! \Gamma(\lambda)}
\int Dx \frac {z^{\mu_1...\mu_n}x^{\mu_1...\mu_n} }{x^{2\gamma}(z-x)^{2\beta}}
\cdot \Biggl[
\frac{1}{\lambda +n+1-\alpha}  \label{11} \\
%\nonumber \\
%\z 
& & \hspace{0.5cm} \times
%&\cdot& \hspace{-0.5cm}
\bigl(
\frac{\Theta(z^2-x^2)}{z^{2(n+\lambda)}x^{2(\alpha -1)}}~+~
\frac{\Theta(x^2-z^2)}{x^{2(n+\lambda)}z^{2(\alpha -1)}} \bigr)
%\nonumber \\  \z
{}~-~
\frac{1}{\lambda +n+\alpha -1} \cdot \bigl(
\frac{\Theta(z^2-x^2)}{z^{2(n+\lambda +\alpha -1)}}~+~
\frac{\Theta(x^2-z^2)}{x^{2(n+\lambda +\alpha -1) }} \bigr)
\Biggr]  \nonumber 
%\label{11}
\end{eqnarray}

After some algebra we have got (see \cite{4}) the result in the form:
$$ C_f[A(\alpha, \beta, \gamma)]~=~ \frac{1}{(4\pi)^{D}}~
\frac{1}{\Gamma(\lambda)}~
\frac{1}{\alpha -1}~
\Bigl[
\overline I ~-~ \tilde I
\Bigr],$$
where
\footnote{We would like to note that the coefficients in 
Eqs.(\ref{13}) and (\ref{18})
are similar to ones (see \cite{FKV})
%which 
appeared in calculations of FD with massive propagators having the mass $m$. 
%The coefficients in front of $(z^2/m^2)^n$
The representation of the results for these diagrams in the form 
$\sum \varphi_n (z^2/m^2)^n$ ($ \varphi_n $ are the coefficients, which
are similar to ones in Eqs.(\ref{13}) and (\ref{18}))
is very convenient to obtain the results for more complicated FD by 
integration in respect of $m$ (see \cite{DEM1}-\cite{DEM23}) of results less 
complicated FD.}

%$$
\begin{eqnarray}
%\z 
\overline{I}&=&
\sum_{n=0}^{\infty}
\frac{\Gamma(n+2 \lambda )
%a_n(2\lambda)
}{n! \Gamma (2 \lambda) }~
\Biggl[
\frac{1}{\lambda +n+1-\alpha} \cdot \biggl(
A^{n,0}(\alpha -1+\gamma, \beta )+A^{n,0}(n+\lambda +\gamma, \beta )
\biggr)
      \nonumber \\ 
%\z 
&-& \frac{1}{\lambda +n+\alpha -1} \cdot \biggl(
A^{n,0}(\gamma ,\beta )+A^{n,0}(n+\alpha + \lambda +\gamma -1, \beta)
\biggr) \Biggr]  \label{13}\\
%\end{eqnarray}
%\begin{eqnarray}
%\z 
& &     \nonumber \\ 
\tilde{I} &=&
 \frac{ \Gamma(1-\beta)\Gamma(\lambda +1-\alpha)
\Gamma(\lambda -1+  \alpha)\Gamma(1-\beta +\lambda)\Gamma(1- \gamma)
\Gamma(\alpha+\beta+\gamma -\lambda -2)}{\Gamma(2\lambda)
\Gamma(2+\lambda-\alpha -\beta)\Gamma(\alpha +\gamma -1)
\Gamma(2+\lambda-\gamma-\beta)\Gamma(\alpha +\beta -\lambda-1)}
       \nonumber  \\
%\z 
&-&    \sum_{n=0}^{\infty}
\frac{\Gamma(n+2 \lambda )}{n! \Gamma (2 \lambda) }~
\frac{(-1)^n \Gamma(1-\beta)}{\Gamma(\beta -\lambda)} \cdot
\frac{1}{\lambda +n+\alpha -1}   \label{18}  \\
%[
%\z 
&\times&   \Biggl[
 \frac{\Gamma(\alpha+\beta+\gamma -\lambda -2)
\Gamma(2-\alpha  -\gamma )}{\Gamma(3-\alpha -\beta -\gamma -n)
\Gamma(\alpha +\gamma +\lambda-1+n)} ~+~
%                            \nonumber \\ \z
\frac{\Gamma(1-\gamma)\Gamma(\beta +\gamma  -\lambda -1)}
{\Gamma(\gamma-\lambda -n)
\Gamma(2+2\lambda-\beta-\gamma+n)}
%]
\Biggr]  \nonumber
\end{eqnarray}

Thus, a quite simple solution for $A(\alpha, \beta, \gamma)$ is
obtained\footnote{ Before our studies, the possibility to represent
  $C_f[A(\alpha, \beta, \gamma)]$ as a combination of
  $_3F_2$-hypergeometric functions with unit argument, has been
  observed in \cite{Bro}.}. In next section we will consider the
important special case of
 these results.\\

{\bf 2.} {\it As a simple but important example} to apply these
results we consider the diagram $J(1,1,1,1,\alpha )$. It arises in the
framework of a number of calculations (see 
\cite{VPH,Kaz,Vas,Gra,Fadin}).
%Gri}, cite{Va}, \cite{Vas} and- \cite{Gra}). 
Its coefficient
function $I(\alpha )
\equiv C_f[J(1,1,1,1,\alpha )]$ can be found (see \cite{4}) as follows
\begin{eqnarray} 
%\z
I(\alpha )~=
\frac{a_0^4(1)a_0(\alpha )}{a_0(\alpha +2-2\lambda)}~
C_f[J( \lambda,\lambda,\lambda,\lambda,\tilde \alpha )]~~ \mbox{ and }~~
C_f[J( \lambda,\lambda,\lambda,\lambda,\tilde \alpha )] =
C_f[A(\tilde \alpha ,3-\lambda -\tilde \alpha , \lambda)]
% \nonumber\\ \z
\nonumber
\end{eqnarray}
%The latter equation may be obtained by analogy with (\ref{10}).

{}From Eqs. (\ref{13}) and (\ref{18}) we obtain
   \begin{eqnarray}
%\z 
& & I(\alpha )~=~ - \frac{2}{(4\pi)^D}
 \frac{ \Gamma ^2(\lambda)\Gamma(\lambda - \alpha )
\Gamma(\alpha +1 -2 \lambda )}{\Gamma(2\lambda)
\Gamma(3\lambda -\alpha  -1)}
       \label{4.1}  \\
%\z 
& & \hspace{0.5cm} 
\times \Biggl[
\frac{ \Gamma ^2(1/2)\Gamma(3\lambda - \alpha -1)\Gamma(2\lambda - \alpha )
\Gamma(\alpha +1 -2\lambda )}{\Gamma(\lambda)
\Gamma(2\lambda +1/2 -\alpha )\Gamma(1/2-2\lambda +\alpha )}
{}~+~   \sum_{n=0}^{\infty}
\frac{\Gamma(n+2 \lambda )}{ \Gamma (n+\alpha +1) }~
\frac{1}{n+1 -\lambda +\alpha }
\Biggl]   \nonumber
\end{eqnarray}
Note that in \cite{Kaz} Kazakov has got another result for $I(\alpha )$:
   \begin{eqnarray}
%\z 
& &I(\alpha )~=~ - \frac{2}{(4\pi)^D}
 \frac{ \Gamma ^2(\lambda)\Gamma(1-\lambda)\Gamma(\lambda - \alpha )
\Gamma(\alpha +1 -2 \lambda )}{\Gamma(2\lambda)\Gamma(\alpha )
\Gamma(3\lambda -\alpha  -1)}
       \label{4.2}  \\
%\z 
& & 
%\hspace{0.5cm} 
\times \Biggl[
\frac{ \Gamma(\lambda)\Gamma(2- \lambda)\Gamma(\alpha )
\Gamma(3\lambda -\alpha  -1)}{
\Gamma(2\lambda -1)\Gamma(3-2\lambda )}
{}~-~   \sum_{n=0}^{\infty} (-)^{n}
\frac{\Gamma(n+2 \lambda )}{ \Gamma (n+2-\lambda) }~ \biggl(
\frac{1}{n+1 -\lambda +\alpha } ~+~ \frac{1}{n+2\lambda -\alpha } \biggr)
\Biggl]   \nonumber
\end{eqnarray}

{}From Eqs. (\ref{4.1}) and (\ref{4.2}) we obtain the
transformation rule for $_3F_2$-hypergeometric function with argument $-1$:
\begin{eqnarray}
%\z
& & {}_3F_2(2a,b,1;b+1,2-a;-1)~=~b \cdot
\frac{\Gamma(2-a)\Gamma(b+a-1)\Gamma(b-a)
\Gamma(1+a-b)}
{\Gamma(2a)\Gamma(1+b-2a)}
\label{4.3}  \\
%\z 
&-& \frac{1-a}{b+a-1} \cdot {}_3F_2(2a,b,1;b+1,b+a;1)
-~ \frac{b}{1+a-b} \cdot {}_3F_2(2a,1+a-b,1;2+a-b,2-a;-1),
  \nonumber
\end{eqnarray}
where $a=\lambda$ and $b=1-\lambda + \alpha $ are used.

Equation (\ref{4.3}) has been explicitly checked at $a=1$ and $b=2-a$
(i.e. $\lambda =1$ and $\alpha =1$), where the
$_3F_2$-hypergeometric functions may be calculated
exactly. 
%We cannot directly 
It is very difficult to prove Eq.(\ref{4.3}) at arbitrary $a$ and $b$
values: the general proof seems to be non-trivial.
Note that it is different from the equations of \cite{5,PBM} and
may be considered
as a new transformation rule.

\section{Applications}

The above results have been used for evaluation
%calculations 
of very complicated FD
which contribute mostly in calculations based on
%in 
various type of $1/N$ expansions: 
%to the following results:
\begin{itemize}
\item
In the calculation (in \cite{QED3}) of the next-to-leading (NLO)
corrections to the value of dynamical mass generation (see \cite{Nash})
in the framework of three-dimensional Quantum Electrodynamics. 
\item
In the evaluation (in \cite{KoKo}) of the correct value of 
of the leading order contribution to the $\beta$-function
of the $\theta$-term in Chern-Simons theory. The $\beta$-function
is zero in the framework of usual perturbation theory but it takes nonzero
values in $1/N$ expansion (see \cite{korea}).
\item
In the evaluation (in \cite{Fadin}) of NLO corrections to the value of
gluon Regge trajectory (see discussions in \cite{Fadin} and references 
therein).
\item
In the calculation (in \cite{KoLi}) of the next-to-leading 
corrections to the BFKL intercept of spin-dependent part of high-energy
asymptotics of hadron-hadron cross-sections. 
\item
In the calculation (in \cite{KoLi}) of the next-to-leading 
corrections to the BFKL equation at arbitrary conformal spin.
\item
In the evaluation (in \cite{BrKo}) of the 
most complicated parts of ${\rm O}(1/N^3)$ contributions to 
critical exponents of $\phi^4$-theory,
for any spacetime dimensionality $D$.
\end{itemize}

We consider here only basic  steps of the last analysis \cite{BrKo}.
Since the pioneering work of the St Petersburg group~\cite{VPH,Imu},
exploiting conformal invariance~\cite{AMP} of critical phenomena,
it was known that the ${\rm O}(1/N^3)$ terms 
%$\eta_3$
in the large-$N$ critical exponents 
%$\eta$ 
of the non-linear $\sigma$-model,
or equivalently $\phi^4$-theory, in any number $D$ of
spacetime dimensions, derives its maximal complexity
from a single Feynman integral $I(\lambda)$ (see \cite{Imu}):
%The definition of $I$ is~\cite{Imu}
\begin{equation}
I(\lambda)=\left.\frac{\rd}{\rd\De}\ln\Pi(\la,\De)\right|_{\De=0},
\label{def}
\end{equation}
where
\begin{equation}
\Pi(\la,\De)=\frac{{x^{2(\la+\De)}}}{\pi^D}\int\int\frac{\rd^D y \rd^D z}
{y^2z^2(x-y)^{2\la}(x-z)^{2\la}(y-z)^{2(\la+\De)}}\label{Pi}
\end{equation}
is a two-loop two-point integral, with three dressed propagators,
made dimensionless by the appropriate power of $x^2$.

The result, obtained by GP technique, is
\begin{equation}
I(\la)=\Psi(1)-\Psi(1-\la)+\frac{\Phi(\la)-\df13\Psi^{\prime\prime}(\la)
-\df{7}{24}\Psi^{\prime\prime}(1)}{\Psi^\prime(1)-\Psi^\prime(\la)}\,,
\label{ans}
\end{equation}
where $\Psi(x)=\Gamma^\prime(x)/\Gamma(x)$ and 
\begin{equation}
\Phi(\la)=4\int^1_0 d x \frac{x^{2\la-1}}{1-x^2}
\left\{{\rm Li}_2(-x)-{\rm Li}_2(-1)\right\}\,
~~~({\rm Li}_2(x)=\sum_{n>0}x^n/ n^2)
,\label{Phi}
\end{equation}
%with  and
%$$.\\

%\vskip 0.5cm

In \cite{BGK,BrKo} the integral $I(\la)$ has been expanded near
$D=2$ and $D=3$, respectively, (i.e. for $D=2-2\varepsilon$ and 
$D=3-2\varepsilon$)
up to $\varepsilon^8$ in the form of alternative and non-alternative
double Euler sums \cite{LE,DZ1}.\\

  {\it Acknowledgments.}
%One of the authors (A.V. K.)
%(A. V. K.)
Author
 would like to express his sincerely thanks to the Organizing
 Committees of the Research Workshop ``Calculations for modern and future
Colliders'' and the XVth International Workshop ``High Energy Physics
and Quantum Field Theory''
and especially to E.E. Boss, V.A. Ilyin, D.I. Kazakov and D.V. Shirkov
 for the kind invitation, 
%and 
the financial support
at  such remarkable Conferences, and 
for fruitful discussions.
Author
was supported in part
by Alexander von Humboldt fellowship.


\begin{thebibliography}{99}

\bibitem{1.1} L. Durand, P.M. Fishbane and L.M. Simmons,
J.Math.Phys. 17 (1976) 1973.
\vspace{-2.5mm}
%
\bibitem{2} K.G. Chetyrkin, A.L. Kataev and F.V. Tkachov,
Nucl.Phys. B174 (1980) 447.
%%CITATION = NUPHA,B174,345;%%
\vspace{-2.5mm}
%
\bibitem{2.1} W. Celmaster and R.J. Gonzalves,
Phys.Rev. D21 (1980) 3112;
%%CITATION = PHRVA,D21,3112;%%
A.Terrano,  Phys.Lett. B93 (1980) 424;
%%CITATION = PHLTA,B93,424;%%
B. Lampe and G. Kramer,
 Phys.Scr. 28 (1986) 585.
%%CITATION = PHSTB,28,585;%%
\vspace{-2.5mm}
%
\bibitem{ChSm} K.G. Chetyrkin and V.A. Smirnov,
Phys.Lett. B144 (1984) 419;
%%CITATION = PHLTA,B144,419;%%
 K.G. Chetyrkin,
in: AIHENP 93,
{\it Proceedings of the 3th International Workshop on Software Engineering, 
Artificial Intelligence and Expert Systems}, ed.by 
K.-H. Becks and D. Perret-Gallix, p. 523.
\vspace{-2.5mm}
%
\bibitem{4} A.V. Kotikov, 
Phys.Lett. B375 (1996) 240.
%%CITATION = HEP-PH 9512270;%%
\vspace{-2.5mm}
%
\bibitem{BrKo}
D.J. Broadhurst and A.V. Kotikov,
Phys.Lett. B441 (1998) 345.
%%CITATION = HEP-TH 9612013;%%
\vspace{-2.5mm}
%
\bibitem{3} D.I. Kazakov and A.V. Kotikov,
Theor.Math.Phys. 73 (1987) 1264.
%%CITATION = TMPHA,73,1264;%%
\vspace{-2.5mm}
\bibitem{3.1}  D.I. Kazakov and A.V. Kotikov,
Nucl.Phys. B307 (1988) 721; 
%%CITATION = NUPHA,B307,721;%%
B345 (1990) 299(E);
%%CITATION = NUPHA,B345,299;%%
\vspace{-2.5mm}
\bibitem{3.2}
 A.V.Kotikov, 
Theor.Math.Phys. 78 (1989) 134.
%%CITATION = TMPHA,78,134;%%
\vspace{-2.5mm}
\bibitem{KazKo} D.I. Kazakov and A.V. Kotikov,
Phys.Lett. B291 (1992) 171.
%%CITATION = PHLTA,B291,171;%%
\vspace{-2.5mm}
%
\bibitem{5} W.N.Bailey, Generalized Hypergeometrical series,
New York, 1972.
\vspace{-2.5mm}
%
\bibitem{VPH}
A.N. Vasil'ev, Yu.M. Pis'mak and J.R. Honkonen,
Theor.Math.Phys. 46 (1981) 104;
%%CITATION = TMPHA,46,104;%%
 47 (1981) 465.
%%CITATION = TMPHA,47,465;%%
\vspace{-2.5mm}
%
\bibitem{FKV} J. Fleischer, A.V. Kotikov, and O.L. Veretin, 
Phys.Lett. B417 (1998) 163;
%%CITATION = HEP-PH 9707492;%%
 Acta Phys.Polon. B29 (1998) 261;
%%CITATION = HEP-PH 9808243;%% 
Nucl.Phys. B547 (1999) 343.
%%CITATION = HEP-PH 9808242;%%
\vspace{-2.5mm}
%
%\begin{thebibliography}{40}
%
\bibitem{DEM1} A.V. Kotikov, 
Phys.Lett. B254 (1991) 158, 
%%CITATION = PHLTA,B254,158;%%
Mod.Phys.Lett. A6 (1991) 677. 
%%CITATION = MPLAE,A6,677;%%
\vspace{-2.5mm}
%
\bibitem{DEM2} A.V. Kotikov, 
Phys.Lett. B259 (1991) 314;
%%CITATION = PHLTA,B259,314;%%
B267 (1991) 123; B295 (1992) 409(E).
%%CITATION = PHLTA,B267,123;%%
\vspace{-2.5mm}
%
\bibitem{DEM23} A.V. Kotikov, 
in: 
{\it Proceedings of the XVth International Workshop 
``High Energy Physics and Quantum Field Theory``}~ (hep-ph/0102178).
%on Software Engineering, Artificial Intelligence and Expert Systems}, 
%ed.by 
%K.-H. Becks and D. Perret-Gallix, p. 523.
\vspace{-2.5mm}
%
\bibitem{Bro} D.Broadhurst and J.A.Gracey, preprint
OUT-4102-46,1993. 
%(unpublished).
\vspace{-2.5mm}
%
\bibitem{Kaz} D.I. Kazakov, Theor.Math.Phys. 62 (1985) 84. 
%%CITATION = TMPHA,62,84;%%
\vspace{-2.5mm}
%
\bibitem{Vas} A.N.Vasil'ev, S.E.Derkachov, N.A.Kivel and A.S.Stepanenko,
  Theor.Math.Phys. 94 (1993) 127;
%%CITATION = TMPHA,94,127;%%
 V.A. Kivel, A.S. Stepanenko and  A.N. Vasil'ev,
Nucl.Phys. B424 (1994) 619;
%%CITATION = HEP-TH 9308073;%%
A.N.Vasil'ev and A.S.Stepanenko,
   Theor.Math.Phys. 94 (1993) 471;
%%CITATION = TMPHA,94,471;%%
 97 (1993) 1349.
%%CITATION = TMPHA,97,1349;%%
\vspace{-2.5mm}
\bibitem{Gra} J.A. Gracey, Phys.Lett. B262 (1991) 49;
%%CITATION = PHLTA,B262,49;%%
Int.J.Mod.Phys. A9 (1994) 567, 
%%CITATION = HEP-TH 9306106;%%
727.
%%CITATION = HEP-TH 9306107;%%
\vspace{-2.5mm}
%
%
\bibitem{Fadin}
V.S. Fadin, R. Fiore and M.I. Kotsky,
Phys.Lett. B387 (1996) 593.
%%CITATION = HEP-PH 9605357;%%
\vspace{-2.5mm}
%
\bibitem{PBM} 
A.P.Prudnikov, Yu.A.Brychkov and O.I.Marichev, 
    Integrals and series, Vol.3, New-York, 1990.
\vspace{-2.5mm}
%
\bibitem{QED3} A.V. Kotikov, JETP Lett. 58 (1993) 711.
\vspace{-2.5mm}
%
\bibitem{Nash} T. Appelquist et al., Phys.Rev. D33 (1986) 3774;
Phys.Rev.Lett. 60 (1988) 2575;
%%CITATION = PRLTA,60,2575;%% 
D. Nash, Phys.Rev. Lett. 62 (1989) 3024.
%%CITATION = PRLTA,62,3024;%%
\vspace{-2.5mm}
%
\bibitem{KoKo} I.N. Kondrashuk and A.V. Kotikov, 
Phys.Rev. D53 (1996) 2260.
%%CITATION = HEP-TH 9603076;%%
\vspace{-2.5mm}
%
\bibitem{korea} S.H. Park, 
Phys.Rev. D45 (1992) R3333.
%%CITATION = PHRVA,D45,3332;%%
\vspace{-2.5mm}
%
%
\bibitem{KoLi}
A.V. Kotikov and L.N. Lipatov, 
Nucl.Phys. B582 (2000) 19. 
%%CITATION = HEP-PH 0004008;%%
\vspace{-2.5mm}
%
%
\bibitem{Imu}
A.N. Vasil'ev, Yu.M. Pis'mak and J.R. Honkonen,
Theor.Math.Phys. 50 (1982) 127;
%%CITATION = TMPHA,50,127;%%
W. Bernrenther and F. Wegner,
 Phys.Rev.Lett. 57 (1986) 1382.
%%CITATION = PRLTA,57,1383;%%
\vspace{-2.5mm}
%
\bibitem{AMP}
A.M. Polyakov,
JETP Lett. 12 (1970) 381;
%%CITATION = JTPLA,12,381;%%
M. D'Eramo, L. Peliti, G. Parisi,
Lett. Nuovo Cim. 2 (1971) 878;
E. S. Fradkin and M. Palchik,
Phys.Rept. 44 (1978) 249.
%%CITATION = PRPLC,44,249;%%
\vspace{-2.5mm}
%
\bibitem{BGK}
D.J. Broadhurst, J.A. Gracey, and D. Kreimer, 
Z.Phys. C75 (1997) 559.
%%CITATION = HEP-TH 9607174;%%
\vspace{-2.5mm}
%
\bibitem{LE}
L. Euler,
Novi Comm.Acad.Sci.Petropol. 20 (1775) 140.
\vspace{-2.5mm}
\bibitem{DZ1}
D. Zagier, in: {\it Proc. First European Congress Math.}, Birkh\"auser,
Boston, 1994, Vol II, pp 497-512;
%
D. Borwein, J.M. Borwein and R. Girgensohn,
Proc.Edin.Math.Soc. 38 (1995) 273;
%
J.M. Borwein, D.A. Bradley and D.J. Broadhurst,
Preprint CECM-96-067, OUT-4102-63
(hep-th/9611004); 
%%CITATION = HEP-TH 9611004;%%
%
J.M Borwein, D.J. Broadhurst, and J. Kamnitzer, Preprint 
OUT-4102-88, CECM-99-13, 1999 (hep-th/0004153);
%%CITATION = HEP-TH 0004153;%%
M.E. Hoffman, The algebra of multiple harmonic series,
preprint from Mathematics Department, U.S. Naval Academy, Annapolis (1996).
%
D.J. Broadhurst, Open University preprint OUT-4102-65, 1996 (hep-th/9612012).
%
%
\end{thebibliography}
\end{document}